\documentclass[aps,prl,reprint,groupedaddress]{revtex4-1}

\usepackage{amsmath}
\usepackage{graphicx}
\usepackage{verbatim}

\begin{document}

\newcommand{\vect}[1]{\mathbf{#1}}

\title{Controlling the Flow of Spin and Charge in Nanoscopic Topological Insulators}

\author{John S. Van Dyke}
\author{Dirk K. Morr}
%\thanks{}
\affiliation{University of Illinois at Chicago, Chicago, IL 60607, USA}

\date{\today}

\begin{abstract}
Controlling the flow of spin and charge currents in topological insulators (TIs) is a crucial requirement for applications in quantum computation and spin electronics.
We demonstrate that such control can be established in nanoscopic two-dimensional TIs by breaking their time reversal symmetry via magnetic defects. This allows for the creation of nearly fully spin-polarized charge currents, and the design of highly tunable spin diodes.  Similar effects can also be realized in mesoscale hybrid structures in which TIs interface with ferro- or antiferromagnets.
\end{abstract}

\pacs{}

\maketitle

Topological insulators represent an exotic state of matter, arising from the coupling of spin and orbital degrees of freedom that is characterized by insulating bulk and gapless edge or surface states \cite{Qi2011aa,Hasan2010aa,Ando2013aa}. The latter are topologically protected from backscattering \cite{Roushan2009aa} and localization \cite{Qi2011aa,Hasan2010aa,Ando2013aa} by defects that preserve the TI's time-reversal symmetry (TRS), and are therefore of great interest for envisioned applications \cite{Yokoyama2009aa,Akhmerov2009aa} in quantum computation \cite{Kane1998aa}  and spin electronics  \cite{Wolf2001aa}. Crucial for the realization of these applications is the creation, control and detection of spin-polarized currents \cite{Qi2011aa,Hasan2010aa,Ando2013aa} in TIs on the sub-100nm scale \cite{Xiu2011aa}. While recent groundbreaking experiments \cite{Li14,Tian14,Tang14,Dan14} have detected spin-polarized currents carried by topological surface states of three dimensional TIs up to room temperature, the creation of highly spin-polarized currents combined with the ability to independently control and manipulate the flow of spin and charge at the nanoscopic scale have remained major obstacles for the use of TI-based devices in spintronics or quantum computation.

\begin{figure}
 \includegraphics[width=8.cm]{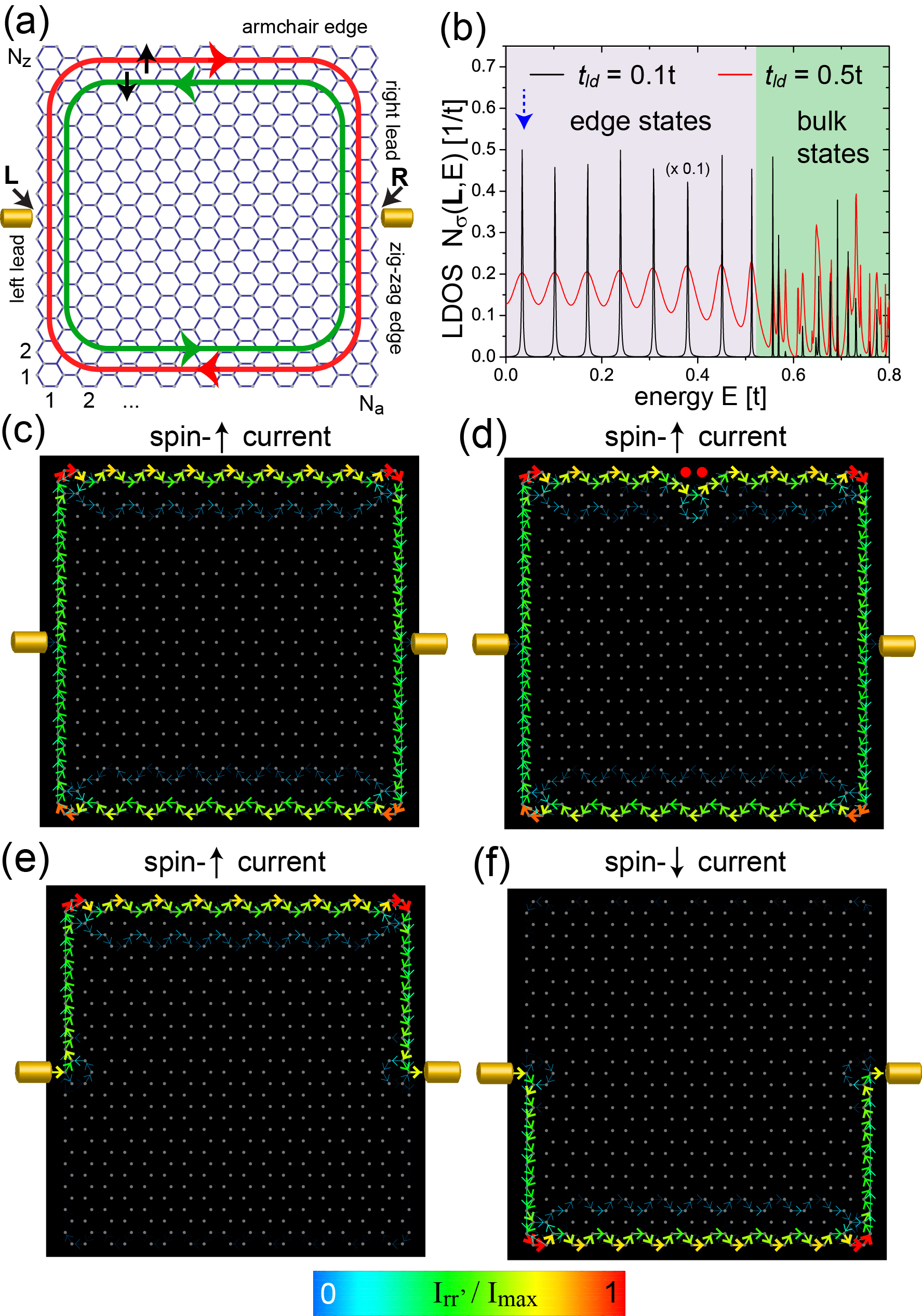}%
 \caption{(a) Schematic representation of the spin-resolved spatial current patterns in a 2D TI with $\Lambda_{SO}=0.1$, $e\Delta V=0.01t$ and $k_B T=10^{-7}t$. (b) Energy dependence of the LDOS, $N_{\sigma}$ at ${\bf L}$ in a clean TI with $N_{\sigma}(-E)=N_{\sigma}(E)$. Spatial pattern of $I_{\vect{r}\vect{r}'}^\uparrow$ for $t_l=0.1t$ carried by (c) the lowest energy edge state at $E_1=0.0342t$ [blue dashed arrow in (b)], and (d) the lowest energy  edge state at $E_1=0.031t$ in the presence of two potential defects (red circles) with scattering strength $U_0=10t$. Spatial pattern of (e) $I_{\vect{r}\vect{r}'}^\uparrow$ and (f) $I_{\vect{r}\vect{r}'}^\downarrow$ carried by the edge state at $E_1=0.0335t$ for $t_l=0.5t$.\label{Fig1}}
 \end{figure}

In this article, we demonstrate that these obstacles can be overcome in nanoscopic two-dimensional (2D) topological insulators \cite{Rasche2013aa,Konig2007aa,Roth2009aa,Young2014aa,Kane2005aa,Xu2006aa,Wu2006aa} by breaking their time-reversal symmetry (TRS) \cite{Chang2014aa,Yu2010aa,Qi2009aa} via magnetic defects \cite{Chen2010aa,Liu2009aa}. In such TI's, the spin-orbit interaction gives rise to the existence of discrete helical edge states \cite{Konig2007aa,Roth2009aa,Young2014aa,Kane2005aa,Xu2006aa,Wu2006aa}, representing Kramers doublets of counter-propagating states with opposite spin polarization [Fig.~\ref{Fig1}(a)]. By placing magnetic defects near the edges of such a TI, one creates spin-polarized currents, either by lifting the degeneracy between the spin-polarized states of a Kramers doublet, or by selectively blocking the flow of current of one spin-polarization while not affecting the other one. It is the superposition of both mechanisms that allows for the emergence of tunable spin-diodes providing control over the flow of charge and spin at the nanoscale. Finally, we demonstrate that similar effects can also be achieved in mesoscale hybrid structures where TIs interface with ferro- or antiferromagnets. These results open new possibilities for the use of TIs in spin-electronics and quantum computation.

	To demonstrate the ability to create and manipulate spin-polarized currents in TIs,  we consider a finite two-dimensional topological insulator [Fig.~\ref{Fig1}(a)] with a hexagonal (graphene-like) lattice structure described by the Kane-Mele Hamiltonian \cite{Kane2005ab}
\begin{align}
H &= -t \sum_{\langle \vect{r},\vect{r}' \rangle,\alpha} c^\dagger_{\vect{r},\alpha}c_{\vect{r}',\alpha} + i\Lambda_{SO} \hspace{-0.3cm}\sum_{\langle \langle \vect{r},\vect{r}'\rangle \rangle,\alpha,\beta} \hspace{-0.1cm} \nu_{\vect{r},\vect{r}'} c^\dagger_{\vect{r},\alpha} \sigma_{\alpha,\beta}^z c_{\vect{r}',\beta} \nonumber \\
& \hspace{1cm} - t_l \sum_{\vect{r},\vect{r}',\alpha} ( d^\dagger_{\vect{r},\alpha} c_{\vect{r}',\alpha} + h.c. ) + H_l \ ,
\end{align}
where the first three terms on the right-hand-side represent the conventional electronic hopping of electrons between nearest-neighbor sites, the intrinsic spin-orbit induced hopping between next-nearest neighbor sites (with $\nu_{\vect{r},\vect{r}'} = -\nu_{\vect{r}',\vect{r}} = \pm 1$, and $\sigma_{\alpha \beta}^z$ being a Pauli matrix) and the hopping between the TI and the leads, respectively, and $H_l$ describes the electronic structure of the leads. We investigate the form of charge and spin transport by employing the non-equilibrium Keldysh Green's function formalism \cite{Keldysh1965aa,Caroli1971aa} where the spin-resolved current between sites $\bf{r}$ and $\bf{r}^\prime$ in the TI is given by
\begin{equation}
I_{\vect{r} \vect{r}'}^\sigma = -2\frac{e}{\hbar} \int _\infty^\infty \frac{\mathrm{d} \omega }{2 \pi} \mathrm{Re} [ t_{\vect{r} \vect{r}'}^\sigma G_{\vect{r} \vect{r}'}^< (\sigma,\omega)] \ .
\end{equation}
Here $\sigma=\uparrow,\downarrow$, $t_{\vect{r} \vect{r}'}^\sigma$ is the real ($-t$) or imaginary ($\pm i \Lambda_{SO}$) electron hopping, and  $G_{\vect{r} \vect{r}'}^< (\sigma,\omega)$ is the full lesser Green's function. The charge current is then given by $I_{out}^c=I_{out}^\uparrow+I_{out}^\downarrow$, and the spin-$\sigma$ polarization of the outgoing current is defined via $\eta_\sigma=I_{out}^\sigma / I_{out}^c$. A current is induced by applying different chemical potentials, $\mu_{L,R}=\pm e\Delta V/2$, in the left ($L$) and right ($R$) leads, resulting in a voltage bias $\Delta V$ across the TI.

	The ability to create highly spin-polarized currents in nanoscopic TIs is largely independent of their particular size, the width of the attached leads, and the strength of the spin-orbit coupling or of the magnetic scattering (see below) \cite{Dyke2015a}. We therefore consider a prototypical nanoscopic TI with $N_a=9$ and $N_z=15$  connected to two narrow leads at sites ${\bf L}$ and ${\bf R}$ [Fig.~\ref{Fig1}(a)]. Due to the TI's finite size, the local density of states (LDOS), $N_\sigma ({\bf r},E)$  [Fig.~\ref{Fig1}(b)] exhibits a set of discrete edge and bulk states located below and above the spin orbit gap $\Delta_{SO}=3\sqrt{3} \Lambda_{SO}$, respectively \cite{Cha11}. To access these states at energy $E_i$ for charge transport, one applies a gate voltage $V_g=E_i/e$ to the TI. In a clean TI, and for weak coupling to the leads ($t_l=0.1t$), the spin-resolved currents carried by the edge states exhibit two counter-propagating, circulating spatial patterns which are confined to the edges of the TI [Fig.~\ref{Fig1}(c)]. The magnitude of these circulating currents can be much larger than the net charge current through the TI, since they possess quantum mechanical backflow branches \cite{Can2012aa}. With increasing $V_g$, the currents extend further into the TI due to an energy dependent decay length of the edge states along the zig-zag edge \cite{Prada2013aa,Cano-Cortes2013aa,Dyke2015a}. Potential scatterers that preserve the TRS, while not leading to backscattering, nevertheless shift the energies of the edge states and modify the current pattern in their vicinity [Fig.~\ref{Fig1}(d)]. Since the coupling to the leads destroys the phase coherence of the edge states and hence breaks the macroscopic time-reversal symmetry of the TI \cite{Qi2011aa}, a larger $t_l=0.5t$ suppresses the currents' backflow branches \cite{Can2012aa} [cf. Figs.~\ref{Fig1}(c) and (e)] resulting in spin-$\uparrow$ and spin-$\downarrow$ currents spatially separated along opposite edges of the TI [Figs.~\ref{Fig1}(e) and (f)]. Spatial imaging of (non-spin-resolved) edge currents was recently reported in HgTe quantum wells \cite{Nowack2013aa}, and could provide unprecedented insight into the TI's electronic structure if attainable with sub-nm resolution \cite{Can2013aa}.

To break the TI's time-reversal symmetry, we place magnetic defects near the edge of the TI at $\vect{R}$, and describe their exchange interaction with the TI's conduction electrons by  \cite{Chen2010aa,Liu2009aa}
\begin{multline}
H_M = \sum_{\vect{R}} J_z S_{\vect{R}}^z ( c^\dagger_{\vect{R},\uparrow} c_{\vect{R},\uparrow} - c^\dagger_{\vect{R},\downarrow} c_{\vect{R},\downarrow} ) \\ + J_{\pm} ( S_{\vect{R}}^+ c^\dagger_{\vect{R},\downarrow} c_{\vect{R},\uparrow} + S_{\vect{R}}^- c^\dagger_{\vect{R},\uparrow} c_{\vect{R},\downarrow} )
\end{multline}
Since the Kondo temperature \cite{Wu2006aa,Maciejko2009aa} can be strongly suppressed either by the absence of edge states near the Fermi energy \cite{Rossi2006aa} [Fig.~\ref{Fig1}(b)], the use of large-spin defects, or by applying local magnetic fields \cite{Rugar2004aa}, while the topological nature of TIs can persist up to room temperature \cite{Dan14}, there exists a sufficiently large temperature range in which we can consider the magnetic defects to be static in nature \cite{Chen2010aa}.

\begin{figure}
 \includegraphics[width=8.5cm]{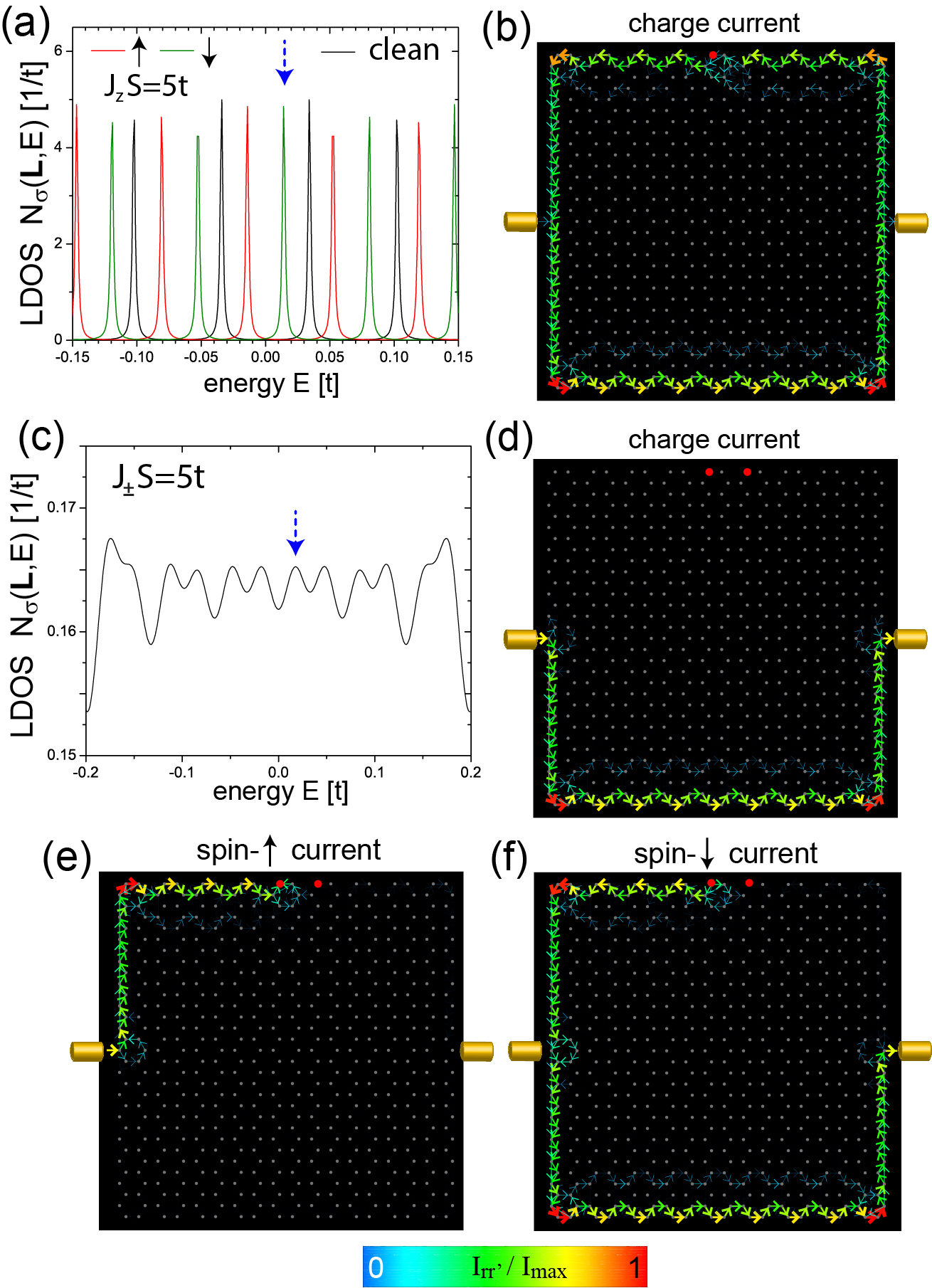}%
 \caption{(a) LDOS, $N_\sigma({\bf L},E)$ in a TI  without and with a magnetic defect [red circle in (b)] with Ising symmetry, $J_z S=5t$ and $t_l=0.1t$. (b) Spatial pattern of the charge current, $I_{\vect{r}\vect{r}'}^c$, carried by the lowest energy edge state at $E_1=0.0142t$ [see blue dashed arrow in (a)]. (c) LDOS, $N_\sigma({\bf L},E)$ for a TI containing two magnetic defects [red circles in (d)] with $xy$-symmetry, $J_\pm S=5t$ and $t_l=0.5t$. Spatial pattern of (d) $I_{\vect{r}\vect{r}'}^c$, (e) $I_{\vect{r}\vect{r}'}^\uparrow$, and (f) $I_{\vect{r}\vect{r}'}^\downarrow$ carried by the edge state at $E_1=0.0175t$  [see blue dashed arrow in (c)] for $t_l=0.5t$. \label{Fig2}}
 \end{figure}

The above interaction enables two qualitatively different mechanisms for the creation of spin-polarized currents via its non-spin-flip ($J_z$) and spin-flip ($J_\pm$)  components. To demonstrate this, we first consider the effect of a magnetic (non-spin-flip) defect with Ising symmetry ($J_z \neq 0$, $J_{\pm}=0$) [Fig.~\ref{Fig2}(b)].
Such a defect lifts the degeneracy of the spin-$\uparrow$ and spin-$\downarrow$ bands by shifting their energies in opposite directions [Fig.~\ref{Fig2}(b)]. When the energy width of these states is smaller than their energy splitting the lifted degeneracy allows one to select a non-degenerate spin-polarized state for current transport via gating. For example, selecting the spin-$\downarrow$ edge state at energy $E_1$ [Fig.~\ref{Fig2}(a)], we find that the charge current is 99\% spin-$\downarrow$ polarized ($\eta_\downarrow=0.99$) with its spatial pattern shown in Fig.~\ref{Fig2}(b). In contrast, when the edge states overlap in energy, either due to a large coupling to the leads, or dephasing arising from an electron-phonon interaction, magnetic (spin-flip) defects of $xy$-symmetry ($J_\pm \neq 0$, $J_z=0$) provide a qualitatively different, but equally efficient mechanism for the creation of spin-polarized currents. These defects scatter electrons between the spin-$\uparrow$ and spin-$\downarrow$ bands, leading to their hybridization, as reflected in the form of the LDOS shown in Fig.~\ref{Fig2}(c). When spin-flip scattering defects are placed into the path of the spin-$\uparrow$ current [Fig.~\ref{Fig2}(e)], they scatter nearly all of the current into the spin-$\downarrow$ band, thus effectively blocking the spin-$\uparrow$ current and creating an additional contribution to the spin-$\downarrow$ current besides the one directly entering from the lead, as shown in Fig.~\ref{Fig2}(f) (a similar effect can occur in the chiral edge states of graphene \cite{Abanin2006aa}). This results in a charge current [Fig.~\ref{Fig2}(d)] with a spin-$\downarrow$ polarization of 96.5\%, which varies only weakly with $V_g$. Finally, we note that for defects with $xy$-symmetry, $\eta_{\sigma}$ remains unchanged under reversal of the gate voltage, while $\eta_{\uparrow} \leftrightarrow \eta_{\downarrow}$ under reversal of the bias $\Delta V \rightarrow -\Delta V$ (and vice versa for defects with Ising symmetry). As a result, under simultaneous reversal of $\Delta V$ and $V_g$, the spin current $i_s=\eta_\uparrow -\eta_\downarrow$ changes sign for both spin symmetries.

\begin{figure}
\includegraphics[width=8cm]{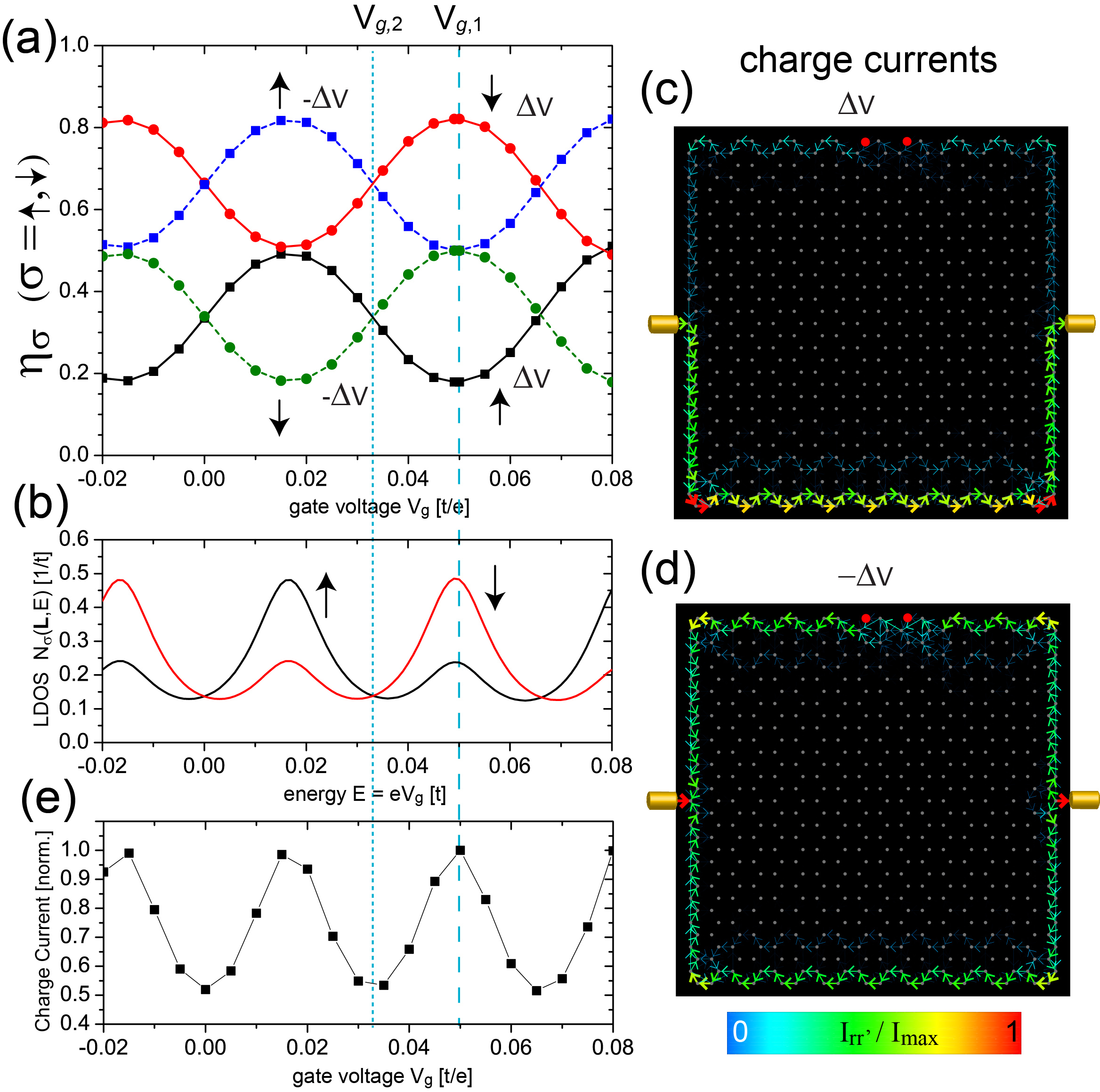}%
\caption{TI containing two magnetic defects of Heisenberg symmetry with $J_z S= J_\pm S=5t$ [red circles in (c)] and $t_l=0.275t$: (a) $\eta_{\uparrow,\downarrow}$ as a function of $V_g$ for forward, $\Delta V$ ($\eta_\uparrow$: black line, $\eta_\downarrow$: red line), and backward bias, $-\Delta V$ ($\eta_\uparrow$: blue dashed line, $\eta_\downarrow$: green dashed line).(b) $N_\sigma(E=eV_g)$ at ${\bf L}$.  Spatial pattern of $I_{\vect{r}\vect{r}'}^c$, for (c) forward bias $\Delta V$, and (d) backward bias $-\Delta V$ at $V_{g,1}$. (e) Total normalized charge current $I^c_{out}(V_g)/I_c^{max}$ with $I_c^{max}=max_{V_g} (I_c^{out})$.}
 \label{Fig3}
\end{figure}
	When isotropic magnetic defects of Heisenberg symmetry ($J_z =J_\pm \neq 0$) are placed into nanoscopic TIs [Fig.~\ref{Fig3}(c)], they lead to a superposition of the two mechanisms discussed above. This allows for the design of highly versatile spin diodes whose characteristics can be tuned via the gate and bias voltages. When $V_g$ is tuned to $V_{g,1}$ [Fig. \ref{Fig3}(a)], the difference between $\eta_\uparrow$ and $\eta_\downarrow$, and hence the spin-current, $i_s=\eta_\uparrow -\eta_\downarrow$, is at a maximum for forward bias, $\Delta V$, but approximately zero for backward bias, $-\Delta V$.  In the latter case, it is the spin-$\downarrow$ current that is scattered by the magnetic defects. However, the resulting loss of spin-$\downarrow$ current is compensated by a larger spin-$\downarrow$ current entering the TI, as expected from $N_\downarrow ({\bf L}) > N_\uparrow ({\bf L})$ at $E=eV_{g,1}$ [Fig.~\ref{Fig3}(b)], thus yielding $\eta_\uparrow \approx \eta_\downarrow$.  At the same time, the charge current remains unchanged under bias reversal. The real space pattern of the charge current, $I_{\vect{r} \vect{r}'}^c$, which can be experimentally imaged \cite{Nowack2013aa,Can2013aa}, directly reflects the extent of the spin current: the large value of $i_s$ for forward bias is reflected in an asymmetric spatial pattern of $I_{\vect{r} \vect{r}'}^c$ [Fig.~\ref{Fig3}(c)] while $i_s \approx 0$ for backward bias implies almost equal currents along the upper and lower edges of the TI [Fig.~\ref{Fig3}(d)]. In contrast, when the gate voltage is tuned to $V_{g,2}$, the spin current $i_s$ changes sign under bias reversal, but its magnitude remains unchanged since $N_\uparrow=N_\downarrow$ [Fig.~\ref{Fig3}(b)]. Moreover, the energy dependence of $N_\sigma ({\bf L})$ implies a change in the charge current with $V_g$ [Fig.~\ref{Fig3}(e)], which is at a maximum for $V_{g,1}$ and at a minimum for $V_{g,2}$. The realization of such TIs therefore provides an intriguing opportunity to design and control spin and charge currents through the applied bias and gate voltages.

\begin{figure}
\includegraphics[width=8.5cm]{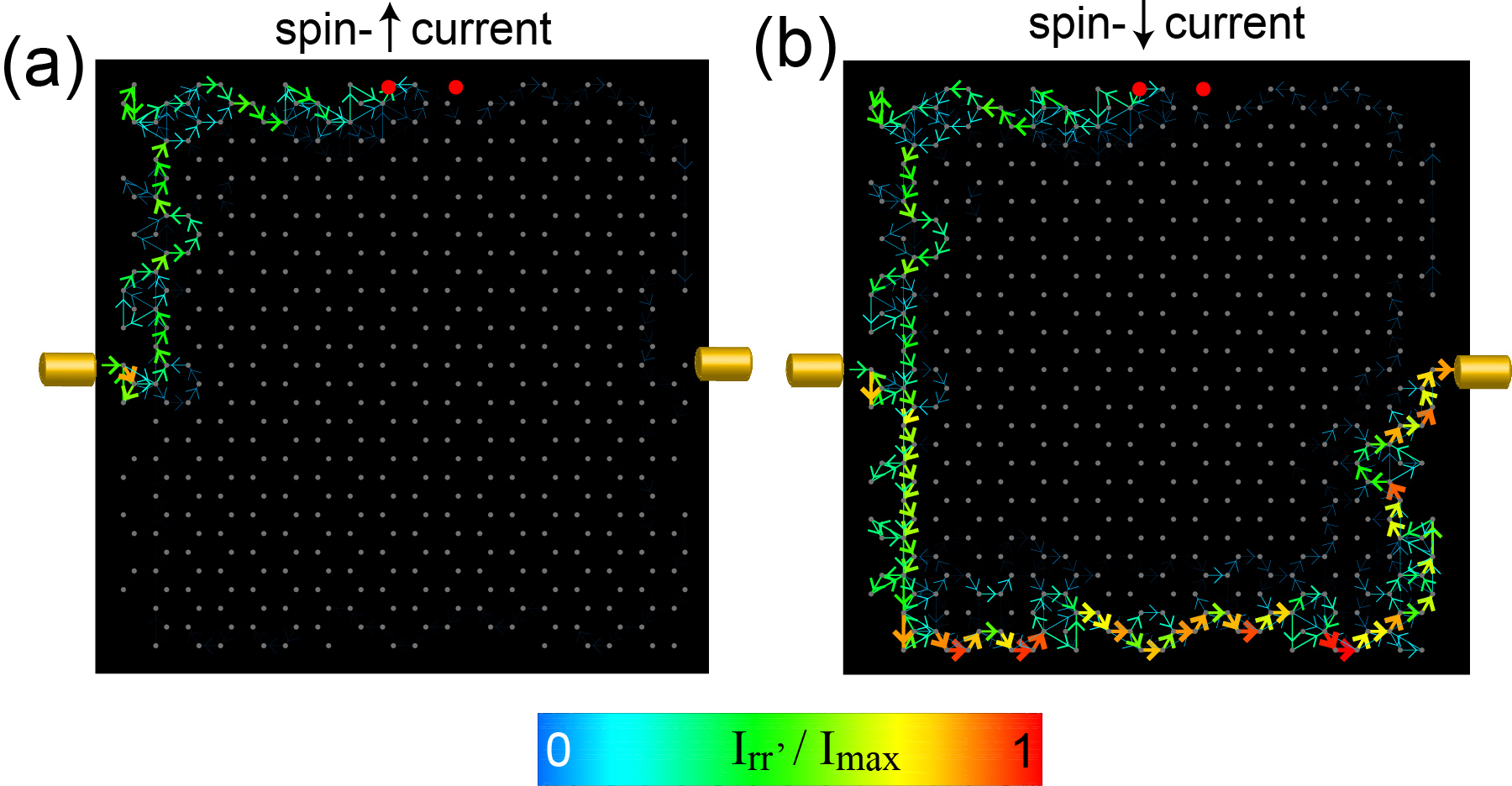}%
\caption{TI with edge disorder and $t_l=0.5t$ and two magnetic defects (red circles) with $J_\pm S=5t$, $J_z S=0$. Spatial pattern of (a) $I_{\vect{r}\vect{r}'}^\uparrow$ and (b) $I_{\vect{r}\vect{r}'}^\downarrow$ carried by the edge state at $E_1=0.04t$. \label{Fig4}}
\end{figure}

The effects discussed above are robust against potential edge disorder, dephasing arising from an electron-phonon interaction, or variations in the width of the leads. Consider, for example, an edge-disordered TI in which 30\% of edge sites are randomly removed (Fig.~\ref{Fig4}) containing two magnetic defects of $xy$-symmetry. While in such a TI, the spatial patterns of the spin-$\uparrow$ and spin-$\downarrow$ currents are more disordered [Figs.~\ref{Fig4}(a) and (b)], the maximum spin polarization (as a function of $V_g$) of $\eta_\downarrow=0.975$ is similar to that of the non-disordered TI [Fig.~\ref{Fig2}(f)] where $\eta_\downarrow=0.965$.  The same robustness of the currents' spin polarization is also found in edge-disordered TIs with Ising-type magnetic defects, or in TIs connected to wide leads (not shown) \cite{Dyke2015a}. Moreover, the interaction with phonons \cite{Bihary2005aa} gives rise to electronic dephasing and a finite scattering time $\tau$. In TIs with magnetic defects of Ising symmetry, the current's spin-polarization is reduced only for $\tau<\hbar/\Delta E$ with $\Delta E$ being the defect-induced energy splitting between the spin-$\uparrow$ and spin-$\downarrow$ edge states. In contrast, in TIs with magnetic defects of $xy$-symmetry, the current's spin polarization is only lost once $\tau$ becomes shorter than $\hbar/\Delta_{SO}$, and the edge and bulk states begin to hybridize \cite{Dyke2015a}. However, in this limit, the topological nature of the insulator is destroyed as well. Similarly, the effects are not tied to particular values of the magnetic scattering strength or the spin-orbit coupling. For example, for the TI shown in Fig.~\ref{Fig2}(e),(f), $\eta_\downarrow=0.96$ decreases only for $J_\pm S \lesssim 1.5t$ (not shown). However, in this case, a high spin-polarization can be restored by increasing the number of defects. Similarly, reducing the strength of the spin orbit coupling to $\Lambda_{SO} = 0.05t$ leads only to a small reduction in the spin polarization to $\eta_\downarrow=0.95$ \cite{Dyke2015a}.

\begin{figure}
\includegraphics[width=8.5cm]{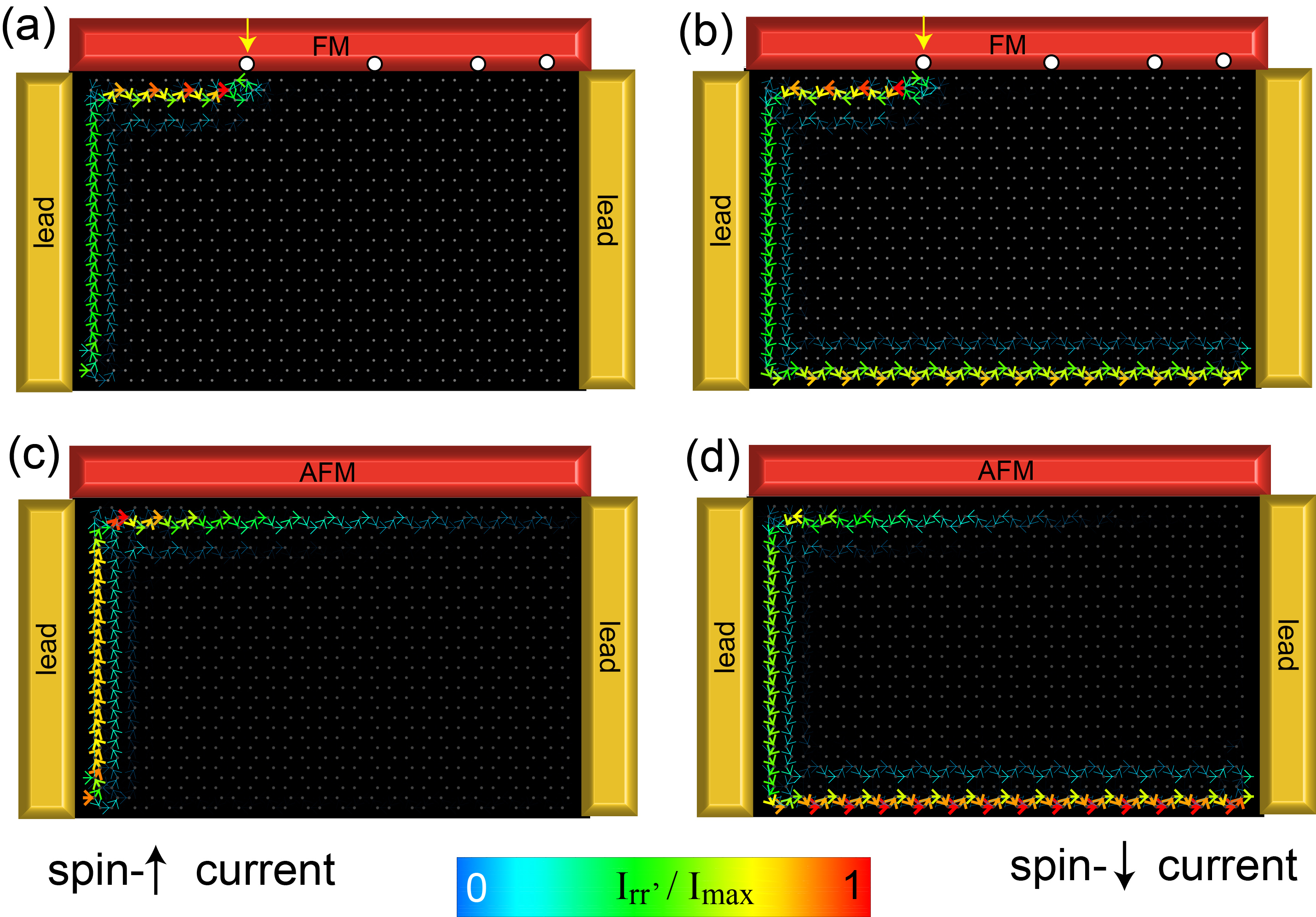}%
\caption{Hybrid structure of TI ($N_a=14$, $N_z=15$, $t_l=0.5t$) and (a),(b) a disordered ferromagnet with $J_\pm S=5t$, the sites with $J_\pm S=0$ are indicated by white circles, and (c),(d) an antiferromagnet with $J_\pm S=\pm 5t$ with an alternating sign between neighboring sites. Spatial pattern of (a),(c) $I_{\vect{r}\vect{r}'}^\uparrow$, and (b),(d) $I_{\vect{r}\vect{r}'}^\downarrow$ carried by the edge state at $E_1=0.1t$ \label{Fig5}}
\end{figure}
 A proof of concept for the effects described above can be achieved in meso- or macroscopic hybrid structures consisting of topological insulators and magnets. Consider, for example, the case of a TI interfacing along one of its edges with an easy-plane ferromagnet (FM) of $xy$-symmetry [Fig.~\ref{Fig5}(a),(b)]. Since in any experimental realization of such a system, it is very likely that the scattering of the TI's edge state electrons off the FM is disordered, we assume random vacancies along the interface (indicated by white circles) where the magnetic scattering vanishes, yielding a spin-polarization of $\eta_\downarrow=0.99$.  Interestingly enough, replacing the ferromagnet by an antiferromagnet (where the sign of $J_\pm S$ changes between neighboring sites), also yields a highly spin-polarized current with $\eta_\downarrow=0.90$ [Fig.~\ref{Fig5}(c),(d)]. Since each pair of neighboring sites (with antiferromagnetically aligned spin) leads by itself to a small spin-polarization,  $\eta_\downarrow$ increases with increasing length of the TI's edge to $\eta_\downarrow=0.96$ for $N_a=18$ and to $\eta_\downarrow=0.99$ for $N_a=25$.  As a result, in the presence of an AFM, scattering of electrons between the spin-$\uparrow$ and spin-$\downarrow$ bands occurs more gradually along the edge [Fig.~\ref{Fig5}(c),(d)], while for the FM cases, the scattering occurs predominantly near the first site where a strong variation in $J_\pm S$ occurs [see yellow arrows in Figs. \ref{Fig5} (a),(b)], Since $\eta_\downarrow$ remains large when the size of these hybrid structures is further increased, we conclude that the ability to create highly spin-polarized currents will likely persist to the meso- and macroscale.

Finally, we find that the effects discussed above are robust against the inclusion of more complicated spin-orbit interactions. In particular, while the inclusion of a Rashba spin-orbit interaction can, for sufficiently large interaction strength $\Lambda_R$ destroy the topological nature of the system, and hence the effects discussed above \cite{Kane2005aa}, we find that even for $\Lambda_R = \Lambda_{SO}$, the currents' spin-polarization $\eta$ is only weakly suppressed by 3-4\% from the values shown above.

In summary, we demonstrated that by breaking the time-reversal symmetry of nanoscopic TIs via magnetic defects, one can create nearly fully spin-polarized currents and design tunable spin diodes. These phenomena are not tied to a particular strength of the defect scattering, are robust against the presence of edge disorder, and persist even in the presence of dephasing, as long as the topological nature of system is not destroyed. As such, they represent universal features of nanoscopic topological insulators that might find use in a wide range of applications from spin-electronics to quantum computation.  We speculate that by employing more elaborate spatial arrangements of magnetic defects, one might be able to further optimize the maximum attainable spin current and tunability of the spin diode, thus allowing for a more independent control of charge and spin currents.

\begin{acknowledgments}
We would like to thank P. Lee, J. Van Den Brink and M. Vojta for helpful discussions.  This work was supported by the U. S. Department of Energy, Office of Science, Basic Energy Sciences, under Award No. DE-FG02-05ER46225.
\end{acknowledgments}

% FIGURES


\end{document}